\documentclass[USenglish,oneside,twocolumn]{article}

\usepackage[utf8]{inputenc}
\usepackage[big]{dgruyter_NEW}
\usepackage{booktabs}
 
\DOI{foobar}

\cclogo{\includegraphics{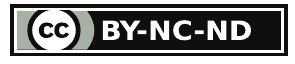}}
  
\begin{document}

  \author[1]{Erick Galinkin}

  \affil[1]{Drexel University, E-mail: eg657@drexel.edu}

  \title{\huge The Influence of Dropout on Membership Inference in Differentially Private Models}

  \runningtitle{Influence of Dropout on Membership Inference}


  \begin{abstract}
{Differentially private models seek to protect the privacy of data the model is trained on, making it an important component of model security and privacy.
At the same time, data scientists and machine learning engineers seek to use uncertainty quantification methods to ensure models are as useful and actionable as possible.
We explore the tension between uncertainty quantification via dropout and privacy by conducting membership inference attacks against models with and without differential privacy.
We find that models with large dropout slightly increases a model's risk to succumbing to membership inference attacks in all cases including in differentially private models.}
\end{abstract}

  \journalname{}
\DOI{Editor to enter DOI}
  \startpage{1}
  \received{..}
  \revised{..}
  \accepted{..}

  \journalyear{..}
  \journalvolume{..}
  \journalissue{..}

\maketitle
\section{Introduction}
Much of the work in differential privacy on deep learning focuses on the impact of privacy-preserving techniques on attacks against them and maximizing accuracy on benchmark datasets.
An often overlooked confounding variable in these studies is the interplay between the techniques which enhance privacy and those techniques which are commonly employed to ensure that models are accurate and avoid overfitting.
In particular, we are forced to reckon with the ways in which regularization and uncertainty management techniques impact the efficacy of membership inference attacks against models with and without differential privacy guarantees.

In this work, we specifically deal with the interaction between dropout, differential privacy, and model inference attacks.
Dropout is a commonly-used regularization technique developed by Srivastava~\textit{et al.}~\cite{srivastava2014dropout} that is employed to alleviate overfitting.
Research by Gal and Ghahramani~\cite{gal2016dropout} demonstrates that dropout not only provides valuable regularization properties, but also gives a meaningful way to make our models more Bayesian and quantify uncertainty.
This is accomplished by incorporating dropout before each and every weight layer, we are providing a ``spike and slab'' prior distribution~\cite{Titsias2011Spike} which makes our model more Bayesian.
This allows us to represent model uncertainty in a meaningful way, which is important since neural networks are notoriously overconfident on data which does not fit its distributions~\cite{kristiadi2020being}.

Once we are convinced the value of uncertainty quantification in predictions and ways we are using, we must also consider how this uncertainty can leak information about members of the training dataset.
In theory, when making a prediction, we would expect some meaningful amount of uncertainty unless the network has seen the exact input before and has ``memorized'' that input.
This reflects the work of Carlini \textit{et al.}~\cite{carlini2020extracting} on memorization in large language models and the ability to extract training data from these language models. 
Carlini \textit{et al}.'s work on language models is an extension of earlier work done by some of the same authors~\cite{carlini2019secret} initially testing for unintended memorization in neural networks.
The crux of this issue is that although we cannot prevent model inversion attacks entirely in all cases, we should still seek to protect the privacy of our training data whenever possible.

One more achievable related attack against machine learning models is that of model inference attacks.
These attacks seek to uncover whether or not a particular example was a member of the model's training set.
This attack is much more feasible in general since it better generalizes to the black-box case and is less computationally intensive.

\section{Related Work}
Membership inference attacks work by examining the predictions output when we provide an input to a model. 
They seek to answer the question: ``was this input in this model's training data set?''
This type of attack was initially demonstrated by Shokri \textit{et al.}~\cite{shokri2017membership} where the attack was demonstrated on three different models and four different datasets. 
Shokri \textit{et al.} concluded that membership inference attacks are effective without any prior knowledge about the target model's training data and only so-called ``black box'' access -- access which does not allow for interrogation of anything but the output of the model.
Salem \textit{et al.}~\cite{salem2019ml} build on this work, reducing the computational load of training multiple shadow models and attack models per class  and instead train only one shadow model as we do in this work.
Our work differs from this prior work by focusing more on the Bayesian methods from Gal and Ghahramani~\cite{gal2016dropout}, which provides more information about the model's prediction.

The use of $\epsilon$ differential privacy has taken off since the seminal 2006 paper by Dwork \textit{et al.}~\cite{dwork2006calibrating} where they consider using noise to ensure privacy in statistical databases. 
The paper presents a framework for computing the amount of noise which must be added to a statistical database to guarantee the privacy of the individuals who provided the data.
$\epsilon$ differential privacy has a popular relaxation, $(\epsilon, \delta)$-differential privacy, again by Dwork \textit{et al.}~\cite{dwork2006our}. 
$(\epsilon, \delta)$-differential privacy allows a small part of the probability distribution where the $\epsilon$ guarantee does not hold.

Work by Abadi \textit{et al.}~\cite{abadi2016deep} to implement deep learning with differential privacy brought these privacy guarantees into the mainstream.
This work yielded a differentially-private stochastic gradient descent optimizer, creating $(\epsilon, \delta)$-differential privacy guarantees by performing gradient clipping and adding noise to the optimization process.
$(\epsilon, \delta)$-differential privacy is employed in neural networks more often than other methods of differential privacy.
This is largely because it is less impactful than the stricter $\epsilon$-differential privacy despite still having a considerable effect on model accuracy~\cite{bagdasaryan2019differential}.

Despite the efficacy of differential privacy in protecting training data, a class of attacks which subverts this privacy guarantee is an active area of research.
These attacks, called membership inference attacks asks the model: ``was this in your training data set?'' and has proven quite effective.
Rahman \textit{et al.}~\cite{rahman2018membership} found that membership inference attacks are moderately effective against differentially private deep learning models given the proper conditions.
The authors found that the value of the $\epsilon$ is a very significant factor in whether or not the membership inference attack is effective and that the membership inference attack is far more effective on the relatively more simple MNIST dataset than on the CIFAR-10 dataset.

\section{Methodology and Data}

\subsection{Methods} \label{sec:methods}
The target model is a modified LeNet-5 architecture~\cite{lecun2015lenet} written in Tensorflow 2~\footnote{Code is available at \texttt{https://github.com/erickgalinkin/dropout\_privacy}}.
We consider the 4 following variants of the model for each dataset:
\begin{enumerate}
	\item ``Vanilla'' LeNet-5
	\item LeNet-5 with Dropout
	\item LeNet-5 trained with DPSGD
	\item LeNet-5 with Dropout trained with DPSGD
\end{enumerate}
For the models with dropout, we use the uncertainty methods described in Gal and Ghahramani~\cite{gal2016dropout} to create the spike and slab prior setting our dropout rate to 0.5.
The target models which use differential privacy are optimized using the default DPSGD optimizer as described by Abadi~\textit{et al.}~\cite{abadi2016deep}.
In this work, we use the default values for norm clipping and noise generation of 1.5 and 1.3 respectively.
For both models, we set the same learning rate of 0.1 to ensure that they are able to be compared directly.

The DPSGD optimizer features a variety of hyperaparameters that can be tuned increase the privacy guarantee and to improve the speed of learning in the model.
There are also a number of hyperparameters in all models that can be tuned and model architectures which could be tried in order to optimize the ratio of test set accuracy to privacy protection.
We reserve these experiments for future work.

The computer vision models are trained for up to 25 epochs.
In order to ensure the best classification accuracy on the test set, we implement early stopping and monitor the test set accuracy, stopping training when test accuracy plateaus.
The trained target models are then saved to disk.

We train a single shadow model using the methods of Salem~\textit{et al.}~\cite{salem2019ml}, building upon the methods of Rahman~\textit{et al.}~\cite{rahman2018membership}.
The shadow model used in this work is a VGG inspired model~\cite{simonyan2015very}.
Our shadow model is then used to train the attack model, which accepts a vector of probabilities from a model output and outputs the probability that the image which produced that probability vector was in the training data set for that model.
The shadow model is used to create the dataset which trains the attack model using the process described in section~\ref{sec:data}.

The attack model is a 2-layer fully-connected neural network.
Given a 10-dimensional input vector, the network learns to predict the probability that the model which output the vector was also trained on that sample.
The model uses the tanh activation in the hidden layers since the more common ReLU activation had worse attack accuracy.

\subsection{Data} \label{sec:data}
In keeping with the previous literature~\cite{rahman2018membership, shokri2017membership}, we use both the CIFAR-10~\cite{krizhevsky2009learning} and MNIST~\cite{lecun1998mnist} datasets to train and evaluate our models.
CIFAR-10 is a well-known benchmark dataset in computer vision, a subset of the now defunct~\cite{birhanelarge} 80 million tiny images dataset. 
The dataset features 10 classes, each with 6000 32x32 images, with 5000 training and 1000 test images per class.
MNIST is another well-known benchmark dataset in computer vision, made up of handwritten digits, 0 through 9. 
The dataset is broken up into 60000 training and 10000 test images with equal representation from all 10 classes.

The dataset for the shadow model is a modified version of the source dataset.
For each image in the source dataset, the image is randomly resized, cropped, flipped, and rescaled to ensure that there is no image in the shadow dataset which is in the source dataset. 
Using this process, however, ensures that the distribution of the source dataset is maintained. 
In order to train the attack model, the source dataset is run through the shadow model to output predictions. 
We then subsample these predictions to create a balanced dataset of in-training and not-in-training predictions coupled with binary labels.
This subsampling is critical to ensuring the accuracy of the attack models, since the datasets in their entirety are heavily imbalanced.

\section{Results}
Although it is not the focus of this work, the probabilities for test-set data were spot-checked.
We found that models trained with the dropout-induced spike-and-slab prior had generally less confident predictions on previously unseen data than the models without dropout for both the differentially-private and non-differentially private case.
This is the essential conclusion of prior work, and our experiment here confirms their finding.

\begin{table*}[ht!]
\centering
\begin{tabular}{@{}l|llll@{}}
\toprule
                            & CIFAR-10 Test    & Attack Efficacy  & MNIST Test       & Attack Efficacy  \\ \midrule
Vanilla LeNet-5             & \textbf{54.41\%} & 54.76\%          & \textbf{97.91\%} & 58.82\%          \\
LeNet-5 with dropout        & 51.04\%          & 60.16\%          & 94.81\%          & 64.60\%          \\
LeNet-5 with DP             & 51.91\%          & \textbf{53.57\%} & 97.62\%          & \textbf{56.74\%} \\
LeNet-5 with DP \& dropout  & 53.26\%          & 60.01\%          & 93.50\%          & 58.62\%          \\ \bottomrule
\end{tabular}
\caption{Summary of results for model test set accuracy and accuracy of attack on the model.} \label{tab:results}
\end{table*}

We summarize the results of the attack experiment in Table~\ref{tab:results}. 
From the results, we observe that there is a meaningful improvement in defense against the attack by using differential privacy.
In particular, the models trained with DPSGD and not dropout was better at defending against model inference on both datasets, with an attack efficacy rate of 53.75\% and 56.74\% versus the worst case of dropout and no differential privacy: 60.16\% and 64.64\%. 
There is, however, a cost associated with using differentially private training in terms of test accuracy, as the baseline LeNet-5 model achieved the best test set accuracy on both datasets and the differentially private models were meaningfully worse on CIFAR-10. 

The effect of dropout can be seen most clearly on the CIFAR-10 dataset, where both with and without differential privacy, the attack was more successful, yielding an attack efficacy rate of 60.16\% and 60.01\% compared with the vanilla and non-dropout DP models achieving only 54.76\% and 53.57\% attack efficacy respectivel.
To determine if this was a significant effect, we establish the null hypothesis that models with and without dropout are equally vulnerable to membership inference attacks.
Using an independent sample t-test across the models given the null hypothesis found that there was significant difference between models with and without dropout $p = 0.015429$.

\begin{figure}[h!]
\centering
\includegraphics[width=0.5\textwidth]{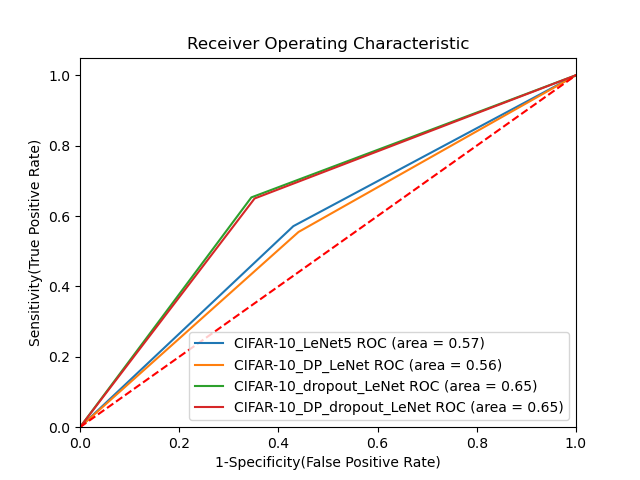}
\caption{Area under the curve plot for CIFAR-10 attack} \label{fig:auc_cifar}
\end{figure}

Figure~\ref{fig:auc_cifar} shows the area under the curve plot for the attack efficacy on the CIFAR-10 dataset. 
We find that the models with dropout have nearly the same AUC in all cases for CIFAR-10, and the differentially private model without dropout has a marginally better curve than the vanilla model.

\begin{figure}[h!]
	\centering
	\includegraphics[width=0.5\textwidth]{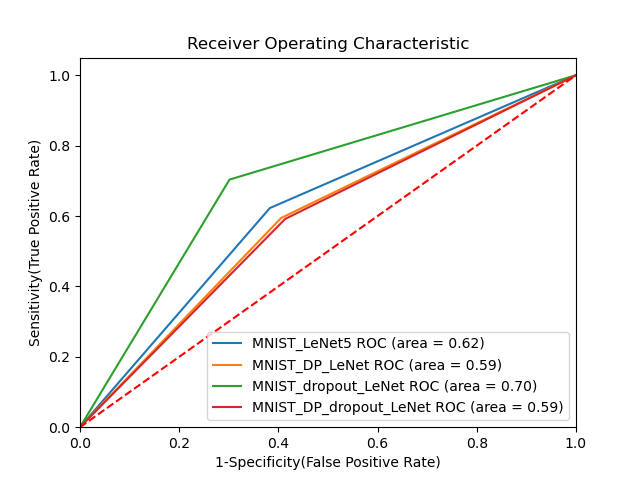}
	\caption{Area under the curve plot for MNIST attack} \label{fig:auc_mnist}
\end{figure}

In Figure~\ref{fig:auc_mnist}, we observe that the green line depicting the AUC for the non-differentially-private model with dropout is substantially higher than the other lines, demonstrating the high efficacy that the attack had on that dataset and model combination.
The differentially private models are much closer together for the MNIST dataset than they are on the CIFAR-10 dataset, suggesting that the dataset is a meaningful factor in attack efficacy.
We note that in both Figure~\ref{fig:auc_cifar} and Figure~\ref{fig:auc_mnist}, the membership inference attack performs better than the baseline 0.5 AUC that random guessing would offer in all cases.

As an informal metric, we consider the ratio of attack efficacy to test set accuracy in Table~\ref{tab:ratio}.
We observe that by this metric, the effects of dropout are more apparent, as is the effect of differential privacy.
We address the ramifications of this in Section~\ref{sec:conclusion}.

\begin{table}[h!]
\centering
\begin{tabular}{@{}l|ll@{}}
\toprule
                            & CIFAR-10 & MNIST  \\ \midrule
Vanilla LeNet-5             & 1.006    & 0.601  \\
LeNet-5 with dropout        & 1.179    & 0.682  \\
LeNet-5 with DP             & 1.032    & 0.581  \\
LeNet-5 with DP \& dropout  & 1.127    & 0.627  \\ \bottomrule
\end{tabular}
\caption{Ratio of attack efficacy to test set accuracy} \label{tab:ratio}
\end{table}

\section{Conclusion and Future Work} \label{sec:conclusion}
We find that likely due to the valuable information provided by moderating the confidence of predictions, dropout as implemented in Gal and Ghahramani~\cite{gal2016dropout} also increases the efficacy of membership inference attacks.
Since uncertainty quantification seeks to tell us \textit{a priori} whether or not a sample is familiar in the way that model inference seeks to derive the information \textit{a posteriori}, it's reasonable to assume that a model that quantifies aleatoric uncertainty will give high uncertainty for new points, and thus be easier to derive information about points which have already been seen.
This means that there is an implicit tradeoff between improving our ability to avoid membership inference attacks and our ability to quantify model uncertainty.

One non-privacy drawback of using dropout before every layer is the creation of up to $d$ additional hyperparameters to tune, where $d$ is the depth of the network.
This can lead to significant training burden when optimizing the level of dropout to use and is likely infeasible for larger networks.
Despite this finding, in cases where the privacy of training set data is not needed, the models trained with dropout had much more reasonable probabilities when making predictions on previously unseen data.
In cases where there is a sensitivity to highly confident predictions, it stands to reason that there is significant value in this approach.

As suggested in Section~\ref{sec:methods}, hyperparameter tuning to find the optimal ratio of attack efficacy to test set accuracy is another fruitful opportunity for future work, as highly accurate models which are also highly resistant to attack is ideal.
This could be done by using a grid search to find the values for the differential privacy mechanism that minimize attack efficacy while also searching for learning rate, early stopping, and dropout or normalization rates that maximize accuracy.
Other model architectures may also have some inherently more or less interesting properties as it relates to privacy.

Our example here dealt only with relatively small and well-studied computer vision datasets, so there is significant opportunity for future work.
As we observed, the dataset itself had an effect on the ability to resist membership inference attacks, with the higher-complexity CIFAR-10 dataset having a better best-case resistance and MNIST having a worse worst-case.
Work on natural language datasets and models such as LSTM and Transformers could provide additional insights into the usefulness of dropout and differential privacy as a defense against membership inference attacks in the general case.
Comparing differential privacy plus dropout against differential privacy plus other Bayesian neural network methods is also a fruitful avenue for follow-up, since the natural question is whether there is something unique about dropout that changes the efficacy of differential privacy against membership inference attacks.

\bibliographystyle{ACM-Reference-Format}
\bibliography{References}

\end{document}